\documentclass[journal]{IEEEtran}

\IEEEoverridecommandlockouts %/thanks
\usepackage{graphicx,latexsym,subfigure,amsmath,epsfig,latexsym,subfigure,amsmath,cite,amssymb}
\usepackage{amsthm,mathrsfs}
\usepackage{graphicx}
\usepackage{epstopdf}
\usepackage{color}
\usepackage{CJK}
\usepackage{array}
\usepackage{booktabs}
\usepackage[ruled,linesnumbered]{algorithm2e}
\usepackage{mathrsfs}%花体设置
\usepackage{bm}%希腊字母加黑
\usepackage{float}%调整图片位置
\usepackage{setspace}%使用间距宏包
\usepackage{color}%使用颜色宏包
\usepackage{subfigure}%插入并排图片
\usepackage{balance}%均衡最后一页内容
\usepackage{subeqnarray}%大括号公式编号
\usepackage{cases}
\begin{document}
%\begin{spacing}{2}
\title{Secure Mobile Crowdsensing with Deep Learning}

\author{\IEEEauthorblockN{\footnotesize Liang Xiao\IEEEauthorrefmark{1}, Donghua Jiang\IEEEauthorrefmark{1}, Dongjin Xu\IEEEauthorrefmark{1}, Ning An\IEEEauthorrefmark{2}}\\
\IEEEauthorblockA{\IEEEauthorrefmark{1}\footnotesize Dept. of Communication Engineering, Xiamen Univ., Xiamen, China. Email: lxiao@xmu.edu.cn}\\
\IEEEauthorblockA{\IEEEauthorrefmark{2} Dept. of Computer and Information, Hefei Univ. of Technology, Hefei, China. Email: ning.g.an@acm.org}\\
%\IEEEauthorblockA{\IEEEauthorrefmark{3}Dept. of Electronics and Information Tech., Sun Yat-sen Univ., Guangzhou, Guangdong Province, China. Email: chenxiang@mail.sysu.edu.cn}\\
%\IEEEauthorblockA{\IEEEauthorrefmark{4}Dept. of Electrical and Computer Engineering, Univ. of Idaho, Moscow, Idaho, USA. Email: mguizani@ieee.org}
}

\maketitle
\begin{abstract}
In order to stimulate secure sensing for Internet of Things (IoT) applications such as healthcare and traffic monitoring, mobile crowdsensing (MCS) systems have to address security threats, such as jamming, spoofing and faked sensing attacks, during both the sensing and the information exchange processes in large-scale dynamic and heterogenous networks. In this article, we investigate secure mobile crowdsensing and present how to use deep learning (DL) methods such as stacked autoencoder (SAE), deep neural network (DNN), and convolutional neural network (CNN) to improve the MCS security approaches including authentication, privacy protection, faked sensing countermeasures, intrusion detection and anti-jamming transmissions in MCS. We discuss the performance gain of these DL-based approaches compared with traditional security schemes and identify the challenges that need to be addressed to implement them in practical MCS systems.
\end{abstract}


\IEEEpeerreviewmaketitle

\section*{Introduction}

%\begin{figure*}[!htbp]
%\centering\includegraphics[width=5.5in,height=3.0in]{F1.eps}\\
%\caption{Threats in mobile edge computing.}\label{fig:threat}
%\end{figure*}
Mobile crowdsensing (MCS) applies embedded
sensors such as gyroscope, accelerometer, microphone and
global positioning system (GPS) in mobile devices such as smartphones to provide location-based services for Internet of Things (IoT) applications including traffic monitoring, healthcare, catering recommendation, location service and social networks \cite{yang2015security}. Participants recruited by an MCS platform monitor the surrounding features and upload the sensing reports to MCS servers that in turn extract interest information from the sensing reports. Security and privacy are critical for MCS systems, as mobile devices are controlled by selfish and autonomous users who can launch insider attacks such as faked sensing attacks and have privacy concerns, and the transmission over radio networks such as 3G/4G, WiFi and vehicular ad-hoc networks (VANETs) are vulnerable to security threats such
as jamming, distributed denial of service attacks
(DoS), spoofing attacks, Sybil attacks, faked sensing attacks and smart attacks \cite{alsheikh2017accuracy}.

In this article, we investigate the security and privacy challenges of mobile crowdsensing, such as sensitive data leakage and faked sensing attacks, and review the MCS security solutions such as authentication,  malware detection and data anonymization to protect user privacy and enhance sensing accuracy in heterogenous dynamic MCS systems. It is challenging to design and implement these security solutions in practical MCS systems due to the difficulty in estimating the mobile user mobility and sensing model in dynamic and heterogenous networks with insider attackers and dynamic network traffic.

Deep learning (DL) techniques have received significant research attentions in speech recognition, computer vision and network security. By applying
DL techniques, the large network state space and extensive training data can be compressed to accelerate the learning speed, optimize the feature extraction, and address the ``high-dimensional disaster"\cite{Han2017Two}. In recent years, the proposed DL-based security schemes such as the privacy protection scheme, the authentication
scheme, and the malware detection scheme exceed the benchmark deterministic schemes \cite{shi2017smart,geyi2017}. In this article, we focus on the MCS security techniques based on deep learning, such as stacked auto-encoder (SAE), deep neural network (DNN), convolutional neural network (CNN), recursive neural network (RNN), deep belief networks (DBN) and deep boltzmann machine (DBM).

We focus on the DL-based authentication, privacy protection, faked sensing countermeasures, intrusion detection, and anti-jamming transmissions in MCS. The performance gain of these DL-based approaches compared
with traditional security schemes is presented, showing that deep learning techniques can accelerate the malware detection speed, enhance authentication accuracy, protect data privacy, reduce the faked sensing attack rate, and resist jamming attacks in sensing report transmissions. We discuss the challenges and identify future directions to implement the DL-based security
approaches in practical MCS systems, such as the real-time security process, accurate evaluation of the MCS utility and the backup security solutions.

This article is organized as follow. In the next section we review the MCS attack models. We then describe
the DL-based authentication, privacy protection, faked sensing countermeasures, intrusion detection, and anti-jamming transmissions. Finally, we identify possible directions of future work.

\begin{figure*}[!htbp]
\centering\includegraphics[width=6in]{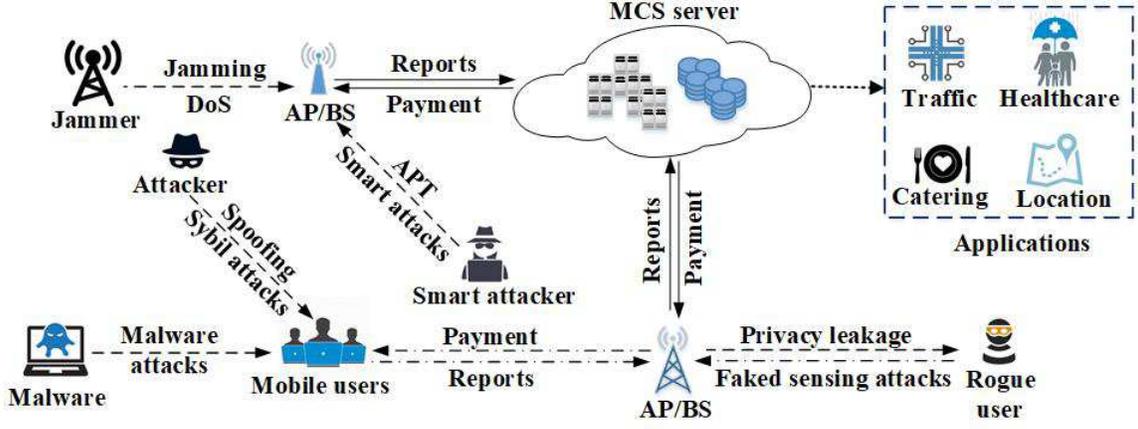}\\
\caption{Threats in mobile crowdsensing systems during the data sensing and information exchange processes.}\label{fig:threat}
\end{figure*}
%\begin{figure*}[!htbp]
%\centering\includegraphics[width=7in]{Threat_MCS_V5.eps}\\
%\caption{Threats in mobile crowdsensing.}\label{fig:threat}
%\end{figure*}
\section*{Threat Model in Mobile Crowdsensing}\label{sec:model}

Malicious attackers in MCS aim to degrade the MCS service levels, crash the MCS system, or steal the privacy data.
MCS systems are also vulnerable to insider attacks launched by selfish mobile users whose goal is to obtain more payments and secrets with less sensing costs and privacy leakage risks. Smart attackers use machine learning techniques and smart radio devices to choose the attack policy and throw more dangerous threats to MCS systems. As illustrated in Fig. \ref{fig:threat}, mobile crowdsensing systems have to address stealthy and insider attacks launched by both selfish mobile users and malicious attackers during the data sensing and information exchange processes to guarantee the veracity and security of multiple MCS applications. We briefly review some significant types of threats in MCS as follows.

%\textbf{Jamming:} A jammer sends faked signals to the edge node or mobile device to interrupt the ongoing radio transmissions. Some jammers also aim to deplete the bandwidth, energy, central processing unit (CPU) and memory resources of the victim edge nodes, mobile devices and sensors by failing their communication attempts \cite{Han2017Two}.

\textbf{Spoofing:} A spoofer uses the identity of another mobile device, a base station (BS) or an access point (AP) to obtain illegal access of the MCS system, and further launch other attacks, such as man-in-the-middle and DoS attacks \cite{shi2017smart}.

\textbf{Sybil:} A Sybil user sends sensing reports with a large
number of different user identities to change the sensing result made by the majority-based MCS server \cite{yang2015security}.

\textbf{Privacy leakage:} MCS servers at the cloud have to protect user privacy especially the user locations and personal information. Rogue users who are curious about the privacy information and steal sensing data from transmission processes or cloud resources, significantly discourage other mobile users from participating in the MCS tasks.

\textbf{Faked sensing attacks:} Selfish mobile users sometimes submit under-sensing or faked sensing reports to save sensing efforts and protect privacy \cite{xiao2017secure}.

\textbf{Malwares:} Malwares such as viruses, worms and spy
tools seriously threaten MCS with the privacy leakage, economic loss, and network performance degradation \cite{geyi2017}. Mobile users with limited power and computing capacity are vulnerable to malwares.

\textbf{Jamming:} A jammer injects faked or replay signals to interrupt the ongoing transmission of the sensing data and payments between mobile users and the MCS servers. A smart jammer can flexibly change the jamming power and frequency according to the transmission status.

\textbf{DoS:} As a typical type of flood attacks, DoS attackers aim to interrupt the MCS services by exhausting MCS server resources\cite{yang2015security}.

\textbf{Advanced persistent threat (APT):} APT attackers launch sophisticated, stealthy, continuous and targeted attacks to steal privacy information from MCS systems over an extended period of time, causing privacy leakage in MCS networks.

\textbf{Smart attacks:} By applying smart radio devices such as universal software radio peripherals (USRPs), an attack can use machine learning techniques to investigate the defense policy and choose attack policy accordingly.

%<{\centering}
\begin{table*}[!htbp]
  \caption{Summary of the DL-based MCS security methods}
\newcommand{\tabincell}[2]{\begin{tabular}{@{}#1@{}}#2\end{tabular}}
  \centering
  \begin{tabular}{p{5.5cm}<{\centering} p{8.5cm} p{2.5cm}} \hline\hline
\specialrule{0em}{3pt}{1pt}
\tabincell{l}{Secure MCS techniques} & Approaches & \tabincell{l}{DL techniques} \\\specialrule{0em}{3pt}{1pt}\hline

\specialrule{0em}{3pt}{1pt}
Authentication & \tabincell{l}{Spoofing detection based on network traffic\\User authentication through daily activities} &\tabincell{l}{SAE \cite{thing2017ieee}\\ DNN \cite{shi2017smart}}  \\\specialrule{0em}{1pt}{3pt}
\hline

\specialrule{0em}{3pt}{1pt}
Privacy protection &\tabincell{l}{Privacy-preserving mechanisms\\Data anonymization\\Image classification based on privacy constraints} & \tabincell{l}{DNN \cite{abadi2016deep}\\CNN \cite{li2017privynet}\\CNN \cite{ossia2017hybrid}}  \\\specialrule{0em}{3pt}{1pt}\hline

\specialrule{0em}{3pt}{1pt}
Faked sensing countermeasure &\tabincell{l}{Payment policy based on sensing accuracy} & \tabincell{l}{DQN \cite{xiao2017secure}} \\\specialrule{0em}{3pt}{1pt}\hline

\specialrule{0em}{3pt}{1pt}
Intrusion detection & \tabincell{l}{Content-based malicious communication detection\\Malware detection with offloading} & \tabincell{l}{RNN \cite{shibahara2016efficient}\\DQN \cite{geyi2017}} \\\specialrule{0em}{3pt}{1pt}\hline

\specialrule{0em}{1pt}{3pt}
Jamming countermeasure & \tabincell{l}{Anti-jamming communication\\Cache-enabled anti-jamming interference alignment} &\tabincell{l}{DQN \cite{Han2017Two}\\DQN \cite{he2017deep}} \\\specialrule{0em}{1pt}{3pt}\hline
\hline
\end{tabular}\label{tabb}
\end{table*}

\section*{DL-based Authentication}\label{sec:authentication}
Authentication verifies the identity of mobile users participated in the MCS tasks to address spoofing and Sybil attacks and protect data privacy for MCS users \cite{kleisouris2010detecting,shi2017smart}. The PHY-layer authentication scheme in \cite{Wang2014Enabling} designs a hypothesis test to compare the radio channel responses of the transmitter under test with the channel record
of the claimed node based on a test threshold that is determined by the given radio channel model and the network model. However, it is challenging to determine the test threshold in time-variant MCS systems with a large number of mobile and heterogeneous devices. Therefore, the DNN-based authentication scheme is developed by \cite{shi2017smart} without requiring the test threshold.

As illustrated in TABLE I, this authentication scheme captures the unique physical features of WiFi signals in the daily activities of mobile users to identify mobile users with the DNN method and develop the support vector machine (SVM) with the generated DNN abstractions to detect spoofing attacks. More specifically, the system extracts 6 time domain features, such as maximum, minimum and skewness, and 3 frequency domain features, including spectrogram energy, percentile frequency component and spectrogram energy difference, from both the amplitude and the phase of the channel responses of WiFi signals. A three-layer DNN based on autoencoder provides the high-level abstractions of human physiological characteristics such as the body shape, height and weight, and the behavioral characteristics such as the walking patterns of the mobile users.

According to the simulation results as shown in \cite{shi2017smart}, the performance of the DNN-based authentication scheme can accurately detect spoofing attacks in typical scenarios. For example, the average accuracy of the scheme is 91.2\% for 7 users with a standard deviation of 3.67\% in the office environment. However, this authentication scheme relies on the tedious and time consuming feature and attribute selection, making it difficult to be applied in practical MCS systems.

To accelerate the MCS authentication process, a spoofing detection system in \cite{thing2017ieee} applies the SAE to capture the features with a neural network (NN) consisting of multiple layers of sparse autoencoders. The first layer learns the first order features from the network traffic, while the following layers learn the features corresponding to the patterns from the previous order features. The first hidden layer in the SAE consists of 256 neurons, the second hidden layer has 128 neurons and the third hidden layer contains 64 neurons. By using the parametric rectified linear unit (ReLU) activation function, the 3-hidden-layer SAE-based authentication scheme achieves accurate and balanced classifications for legitimate mobile users from flooding, injection and spoofing attackers. For example, this SAE-based authentication scheme achieves the spoofing detection accuracy of 98.5\%, which is 14 times higher than the J48-based authentication algorithm with the spoofing detection accuracy of 6.4\%.

\begin{figure*}[!htbp]
\centering\includegraphics[width=6.6in]{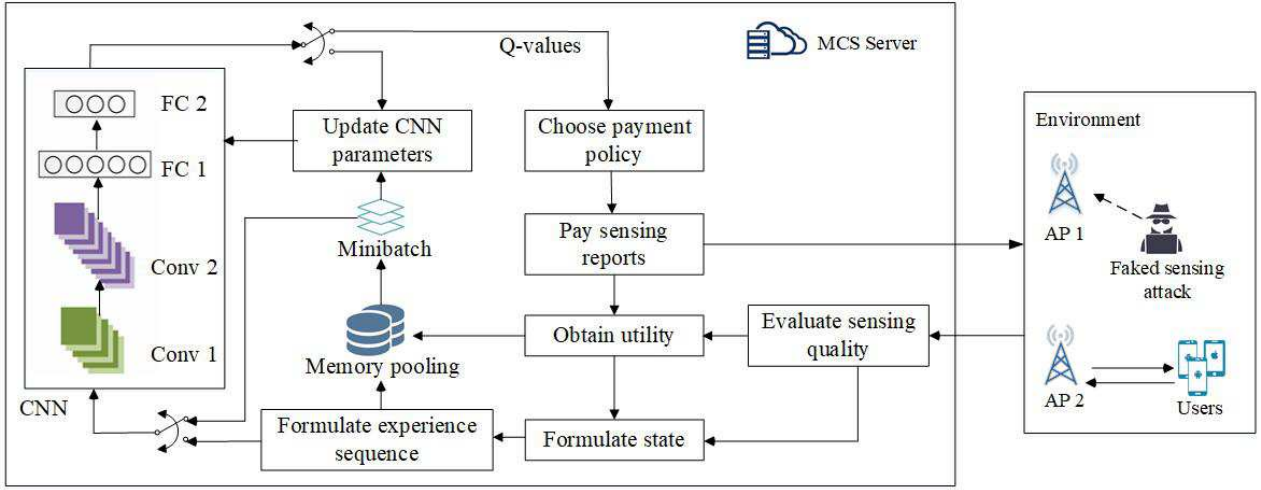}\\
\caption{DQN-based MCS payment scheme to suppress the faked sensing rate with evaluating sensing quality of the mobile users.}\label{fig:DQN}
\end{figure*}

\begin{figure*}[!htbp]
\centering\includegraphics[width=6.7in]{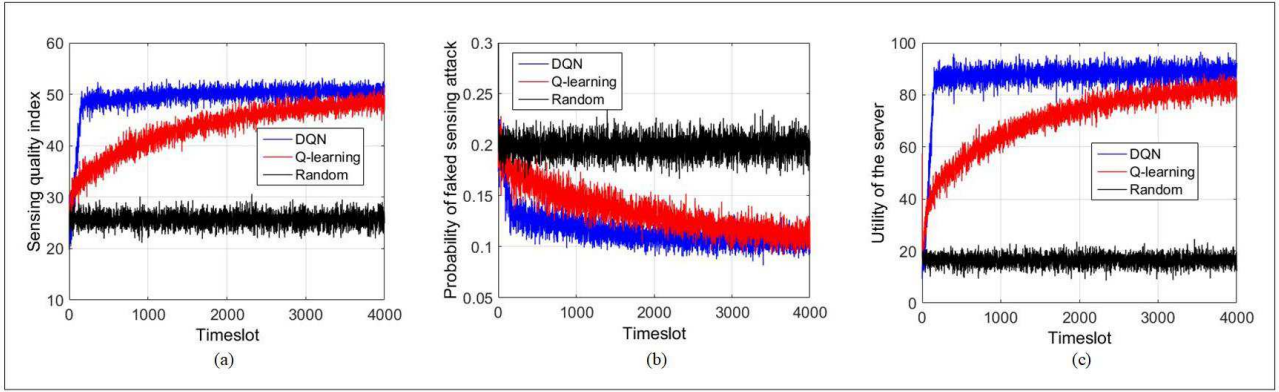}\\
\caption{Performance of the DQN-based dynamic MCS system with 60 mobile users, 2 sensing accuracy levels, 11 feasible payments, and evaluation error rate of the MCS server 0.1.}\label{fig:quality}
\end{figure*}

\section*{DL-based Privacy Protection}\label{sec:privacy}
Privacy protection for sensing report transmission and storage at cloud servers is essential for practical MCS systems. Data anonymization methods, such as $k$-anonymity, $l$-diversity and $t$-closeness \cite{alsheikh2017accuracy}, provide anonymous quasi-identifiers and protect sensitive information in database for MCS. However, the sensing reports containing high-dimensional data streams such as images and videos, cannot be satisfied with such traditional data anonymization schemes, due to the difficulty defining the quasi-identifiers and sensitive attributes.

Therefore, a DNN-based flexible framework as proposed in \cite{li2017privynet} uses DNN training at the mobile devices to provide with the data privacy protection. This scheme consists of the local NN platform that extracts features and the NN at the cloud server that applies the released features to train the given MCS task. The local NN that consists of 4 convolutional layers and 2 max-pooling layers provides data anonymization for MCS systems. The trade-off between user privacy and sensing accuracy depends on the number of DNN layers and the output channel depth. Simulation results provided in \cite{li2017privynet} show that the privacy level and the utility of the DNN-based MCS system decrease with the number of layers and increase with the output depth in this framework.

On the other hand, as the feature extraction network (FEN) in \cite{li2017privynet} depends on the pre-trained NNs in local platform, attackers can predict the structure and weights of the NN in the MSC system. Therefore, a pool of pre-trained NNs enables FEN derivation from different pre-trained NNs to protect the MCS user anonymity. In addition, the channel selection process consisting of the output channels and intermediate channels selection is hard to be known and copied by attackers.

A CNN-based distributed framework as developed in \cite{ossia2017hybrid} uses the siamese network that is widely used by verification applications to optimize the cost function consisting of the data privacy and sensing task accuracy, and applies the CNN to improve the image classification accuracy at MCS servers. In this framework, a feature extractor at the mobile device and image classifier at the MCS server cooperate to perform the MCS tasks, such as emotion detection. Because only convolutional layers are implemented on the mobile devices, the memory overhead significantly decreases at the mobile device and thus both the model initialization time and the MCS power consumption reduce significantly. By removing the undesired information from the extracted features at the mobile device, the framework improves the performance of the MCS tasks, especially gender classification and emotion detection, and prevents the user identity leakage due to the face recognition from the MCS server. For example, the face recognition accuracy of the CNN-based framework with siamese network is 2.6\%, which is 94.5\% lower than that of the framework without siamese network, while the emotion detection accuracy only decreases 20\%.

The DNN-base privacy protection method as proposed in \cite{abadi2016deep} applies different levels of Gaussian noise in the training parameters to provide differential privacy which reduces the sensing accuracy as well. To balance the privacy preservation and the sensing accuracy, a coalition MCS incentive mechanism as designed in \cite{alsheikh2017accuracy} evaluates the user payoff and compares with a threshold that currently is simply set to be zero. However, such a test threshold is not always optimal and sometimes even loses the balance between the privacy protection and sensing accuracy in dynamic MCS networks. Therefore, deep learning techniques such as CNN can be applied by the mediator to achieve the optimal threshold and thus improve the sensing accuracy and user privacy.

\section*{DL-based Faked Sensing Countermeasure}\label{sec:faked}
Trust mechanisms previously designed for VANETs and peer-to-peer networks, such as the anonymous reputation management mechanism as presented in \cite{Wang2014Enabling} are not applicable in MCS with variety of participants and large-scale sensing data \cite{he2015user}, causing long sensing latency and low detection accuracy. To this end, deep learning can reduce the dimension of the sensing data set and accelerate the learning speed of MCS.

As shown in Fig. \ref{fig:DQN}, a DQN-based secure MCS system as proposed in \cite{xiao2017secure} suppresses the faked sensing motivation for autonomous mobile users in dynamic secure mobile crowdsensing games. In this framework, the server first determines and broadcasts its payment policy to mobile users in the area of interests. Each mobile device chooses its sensing effort according to the payment policy and its current battery levels and receives the payment according to the sensing quality evaluated by the MCS server. As the MCS payment process in a dynamic game can be formulated as a Markov decision process, the MCS server can apply reinforcement learning (RL) techniques such as Q-learnig to achieve the optimal MCS policy for the mobile users who might launch faked sensing attacks, and uses a deep CNN that consists of two convolutional (Conv) layers and two fully connected (FC) layers to compress the state space of RL and thus accelerate the learning process. The CNN parameters are updated according to the stochastic gradient descent (SGD), which samples a subset of summand functions at each iteration to reduce the computational cost.

As shown in Fig. \ref{fig:quality}, this MCS payment strategy outperforms the benchmark non-DL scheme with a higher sensing quality, a lower faked sensing rate and a higher utility of the MCS server. For instance, the DQN-based MCS payment scheme only takes 2000 time slots to suppress the faked sensing rate to 10\%, which is 50.0\% faster than the Q-learning based strategy. The proposed payment scheme improves the utility of the MCS server by 96.0\% after 200 time slots, compared with the Q-learning based scheme.
%
%\begin{figure*}[!htbp]
%\centering\includegraphics[width=6.7in]{Faked_sensing_results.eps}\\
%\caption{Performance of the DQN-based MCS system}\label{fig:quality}
%\end{figure*}

\section*{DL-based Intrusion Detection}\label{sec:detection}
In MCS systems, mobile devices with limited power, computing capacity and wireless bandwidth can use cloud-based intrusion detection to detect malwares with fast processing speed, powerful security services and large malware database \cite{geyi2017}.
By applying learning techniques such as Bayes network and random forest classifiers, the MCS system evaluates the runtime behaviors of the applications and processes the traces or logs.
However, the malware detection accuracy is sensitive to the features and attributes, and many existing methods such as the framework proposed in \cite{yang2015security} suffer from a long detection delay in large-scale networks.

\begin{figure}[!htbp]
\centering\includegraphics[width=3.3in]{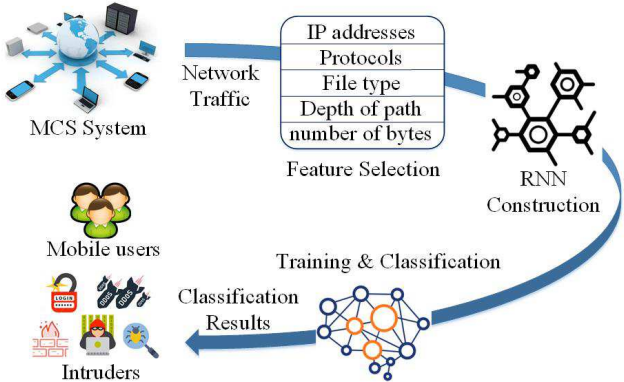}\\
\caption{RNN-based intrusion detection process with 3 steps.}\label{fig:intrusion}
\end{figure}

A RNN-based malware detection scheme as developed in \cite{shibahara2016efficient} uses RNN to capture the domain and content-based malware characteristics and determine whether to suspend the dynamic detection based on the network behavior. As shown in Fig. \ref{fig:intrusion}, the malware detection consists of the feature extraction, RNN construction and training \& classification, based on the domain and content features such as the IP address of the source node, the types of files and the number of bytes. With accurate classification performance especially to process natural languages, RNN can analyze the latent functions of malware communications, such as the malware infection spread and information leakage. Simulation results show that this scheme saves the detection resources and accelerates the response speed compared with the benchmark scheme without deep learning due to the high flexibility of RNN. The RNN-based malware detection outperforms the statistics-based malware detection with 67.1\% less analysis time, reducing the number of MCS servers by more than half for efficient MCS dynamic analysis.

MCS systems can also apply deep learning to optimize the offloading policies without being aware of the radio channel model and the APP traces generation model in malware detection. For instance, a cloud-based mobile malware detection scheme as proposed in \cite{geyi2017} applies DQN to determine the offloading rate of the mobile device. More specifically, the offloading rate is chosen according to the quality function of the offloading strategy and the current state that consists of the current radio bandwidth and the previous offloading rates of the mobile devices.
The Q-function is estimated according to the CNN, which consists of two Conv layers and two FC layers. The CNN parameters are updated according to the SGD method, which minimizes the mean-squared error of the target values via minibatch and the experience replay techniques. Simulation results show that the DQN-based scheme applies the CNN to compress the state space and thus accelerate the learning speed, thus improving the malware detection accuracy and saving the detection time compared with the RL-based benchmark scheme without DL.
For instance, the DQN-based scheme increases the detection accuracy by 24.5\%, reduces the detection delay by 35.3\%, and increases the utility of the mobile device by 31.0\%, compared with the Q-learning based scheme.

\begin{figure*}[!htbp]
\centering\includegraphics[width=4.5in]{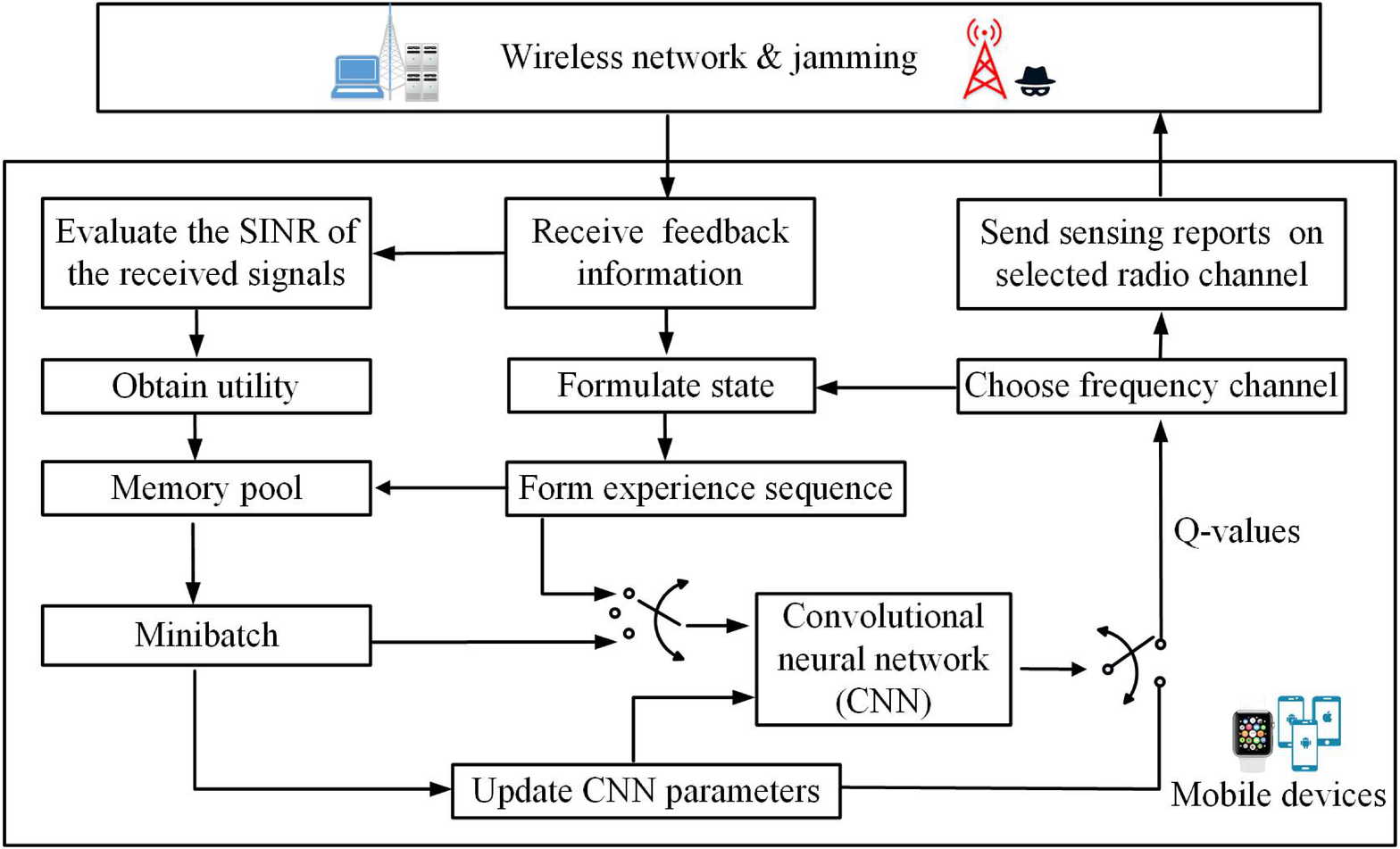}\\
\caption{DQN-based anti-jamming transmissions with 2 smart jammers and 128 radio channels.}\label{fig:jamming}
\end{figure*}
\section*{DL-based Anti-jamming Transmission}\label{sec:anti-jamming}
MCS systems have to address jamming attacks and DoS attacks and improve the transmission efficiency of the sensing data and the payment decisions with high signal-to-noise-plus-interference (SINR), low bit error rates (BERs) and low transmission energy consumptions against jamming. However, traditional anti-jamming techniques such as frequency hopping and direct-sequence spread spectrum are not always applicable in MCS systems with heterogenous and dynamic large-scale networks, causing the loss of the sensing data and payments, which discourages mobile users to participate in the sensing tasks.

A promising solution to address jamming attacks in MCS systems is the DQN-based two-dimensional anti-jamming communication system as proposed in \cite{Han2017Two}. As shown in Fig. \ref{fig:jamming}, this scheme applies frequency hopping to resist jamming and improve the SINR of the received signals. Deep learning techniques such as CNN are used to help such MCS system to balance the sensing \& payment transmission quality and the communication overhead such as the energy consumption and latency of the channel hopping and user mobility. This scheme enables a mobile device to achieve the optimal communication policy according the current transmission state without being aware of the jamming and interference model and the radio channel model in MCS systems. The CNN is used to compress the state space for large scale MCS mobile and heterogeneous devices and thus accelerate the learning speed and increase the SINR of the received signals.

Simulation results as provide in \cite{Han2017Two} show that the DQN-based anti-jamming communication system has a faster convergence rate and higher SINR than Q-learning. For instance, this scheme converges to the optimal performance after 1000 time slots, and saves 66.7\% of the learning time compared with the benchmark scheme. The SINR of the DQN-based communication system is 8.3\% higher than Q-learning after 1000 time slots.
%\begin{figure}[!htbp]
%\centering\includegraphics[width=3.3in]{Anti-jamming_V5.eps}\\
%\caption{DL-based anti-jamming transmission.}\label{fig:jamming}
%\end{figure}

DQN has also been used in a cache-enabled anti-jamming communication scheme as developed in \cite{he2017deep} to optimize the interference alignment and user selection. This scheme provides user cooperation in the design of the precoding matrices to resit jamming. Implemented with Google TensorFlow, this communication systme can work in practical MCS with time-varying radio channels. By utilizing a target CNN network, this anti-jamming communication system periodically updates the values of the Q-network according the outputs of the target network to address the destabilization problem. The learning rate and minibatch size of the CNN are discussed regarding the convergence performance. For example, this scheme has a faster convergence rate with smaller minibatch size, which might increases the risks of local optimum. Therefore, appropriate minibatch size should be chosen properly before being implemented in MCS systems.

\section*{Conclusion \& Future Work}\label{sec:conclusion}
In this paper, we have reviewed the attack models in MCS systems and investigated the DL-based authentication, privacy protection, faked sensing countermeasure, intrusion detection, and anti-jamming transmissions. The DL-based security techniques are promising to improve the security performance of the MCS systems. Some important directions for future study of the DL-based MCS security solutions are suggested below.

%\textbf{Accuracy-privacy trade-off:}
%
%\textbf{General privacy metrics:}
%
%\textbf{Incentive mechanism for fog computing:}

\textbf{Real-time processing:} Most existing deep learning based security schemes such as the DNN-based authentication in \cite{shi2017smart} and the RNN-based malware detection in \cite{shibahara2016efficient} require long training time and are too complicated to be implemented in the practical MCS systems for real-time processing. The widely used hardware for deep learning computation such as graphics processing units (GPUs), is not applicable for most mobile devices such as smartphones in MCS systems. Therefore, distributed deep learning algorithms that distribute the processing and storage tasks among MCS network components and the low-cost DL compatible chips are critical to protect MCS systems in the future. Another interesting topic is to optimize the deep learning architecture and parameters for MCS security solutions, such as the second order stochastic gradient descent that helps accelerate the sensing speed for MCS systems against attacks.

\textbf{Accurate evaluation of utility:} Deep learning based security schemes such as the CNN-based sensing data anonymization in \cite{li2017privynet} and the DQN-based faked sensing countermeasure in \cite{xiao2017secure} have to evaluate the security performance according to the utility function, which depends on the instant security gain and protection cost. For example, an MCS server has to estimate the sensing accuracy, computation cost and protection cost in each time slot to decide the payment policy and suppress the faked sensing motivations of mobile users in the DQN-based MCS system in \cite{xiao2017secure}. However, it is challenging for the MCS system to accurately evaluate such factors and determine the utility in time, especially for the case with stealthy and insider attackers. In the future we have to investigate the deep learning algorithms that are robust against the utility evaluation errors to design the MCS security solutions.

\textbf{Backup security solutions:} Existing DL techniques, such as the DQN-based malware detection with offloading in \cite{geyi2017}, require a long training stage and trial-and-error stage at the beginning, indicating MCS security failures against attacks, which sometimes causes  network disasters and millions of dollars loss. Therefore, backup security protocols have to be incorporated with the DL-based security schemes to provide reliable and secure MCS services.

\ifCLASSOPTIONcaptionsoff
  \newpage
\fi

\bibliography{save_v2}
\bibliographystyle{IEEEtr}

\begin{IEEEbiographynophoto}
%[{\includegraphics[width=1in,height=1.25in,clip,keepaspectratio]{xiao8.eps}}]
{Liang Xiao}
(M'09, SM'13) is currently a Professor in the Department of Communication Engineering, Xiamen University, Fujian, China. She has served in several editorial roles, including an associate editor of IEEE Trans. Information Forensics \& Security and IET Communications. Her research interests include wireless security, smart grids, and wireless communications. She won the best paper award for 2016 IEEE INFOCOM Bigsecurity WS. She received the B.S. degree in communication engineering from Nanjing University of Posts and Telecommunications, China, in 2000, the M.S. degree in electrical engineering from Tsinghua University, China, in 2003, and the Ph.D. degree in electrical engineering from Rutgers University, NJ, in 2009. She was a visiting professor with Princeton University, Virginia Tech, and University of Maryland, College Park. She is a senior member of the IEEE.
\end{IEEEbiographynophoto}

\begin{IEEEbiographynophoto}
%[{\includegraphics[width=1in,height=1.25in,clip,keepaspectratio]{donghua.eps}}]
{Donghua Jiang}
received the B.S. degree in electronic information science and technology from Xiamen University, Xiamen, China, in 2017, where she is currently pursuing the M.S. degree with the Department of Communication Engineering. Her research interests include machine learning, network security and wireless communications.
\end{IEEEbiographynophoto}

\begin{IEEEbiographynophoto}
{Dongjin Xu}
received the B.S. degree in communication engineering from Xiamen University, Xiamen, China, in 2016, where she is currently pursuing the M.S. degree with the Department of Communication Engineering. Her research interests include network security and wireless communications.
\end{IEEEbiographynophoto}

\begin{IEEEbiographynophoto}
{Ning An}
received the B.S. and M.S. degrees in computer science from Lanzhou University, Lanzhou, China, in 1993 and 1996, respectively, and the Ph.D. degree in computer science and technology from Pennsylvania State University, State College, PA,
USA, in 2002. He is a Full Professor in the School of Computer and Information and the Founding Director of the Gerontechnology Laboratory at Hefei University
of Technology, Hefei, China. His research interests include gerontechnology, data science and engineering, Internet of Things, and eHealth.
\end{IEEEbiographynophoto}

%\end{spacing}
\end{document}